\documentclass[preprint,5p]{elsarticle}

\usepackage{amssymb,amsmath}

\newcommand{\gev}{\ \mathrm{GeV}}
\newcommand{\tev}{\ \mathrm{TeV}}
\newcommand{\fb}{\ \mathrm{fb}}
\newcommand{\pb}{\ \mathrm{pb}}

\journal{Physics Letters B}

\begin{document}

\begin{frontmatter}



\title{Early spin determination at the LHC?}
\tnotetext[preprints]{DESY 11-017, Bonn-TH-2011-03}
\author[desy,unih]{Gudrid Moortgat-Pick}
\author[desy]{Krzysztof Rolbiecki\corref{cor1}}
\ead{krzysztof.rolbiecki@desy.de} \cortext[cor1]{Corresponding
author}
\author[bonn]{Jamie Tattersall}
\address[desy]{DESY, Deutsches Elektronen-Synchrotron, Notkestr.\ 85, D-22607 Hamburg, Germany}
\address[unih]{II. Institut f\"{u}r Theoretische Physik, University of Hamburg, Luruper Chaussee 149,
D-22761 Hamburg, Germany}
\address[bonn]{Universit\"{a}t Bonn, Physikalisches Institut, Nu\ss allee 12, 53115 Bonn, Germany}

\begin{abstract}
If signals of new physics are discovered at the LHC it will be
crucial to determine the spin structure of the new model. We discuss a
method that can help to address this question with a low integrated
luminosity, $\mathcal{L}=1\fb^{-1}$, at $\sqrt{s}=14 \tev$. Based on the differences in
angular distributions of primarily produced particles we show that a
significant difference can be observed in the final state jet-pairs
rapidity distance. An additional advantage of the method is that it
does not rely on any particular structure of the couplings in the
decay chain. We simulate samples for models with supersymmetric and
UED-like spin structure and show that a distinction can be made
early on.
\end{abstract}



\end{frontmatter}


\section{Motivation}
\label{sec:intro}

The Large Hadron Collider (LHC) has started operating and will soon
probe physics at the TeV scale, perhaps revealing the origins and
mechanism of the electroweak symmetry breaking. One of the most
promising candidates for explaining this phenomenon is supersymmetry
(SUSY)~\cite{Haber:1984rc,Nilles:1983ge}. In supersymmetric theories
each Standard Model (SM) particle is paired with a superpartner of
spin different by $1/2$. In particular, spin-$1/2$ quarks will be
accompanied by spin-0 squarks, and spin-1 gluons by spin-$1/2$
gluinos. Another possibility is provided by models with universal
extra dimensions (UED)~\cite{Appelquist:2000nn}. In such models each
SM particle will have a tower of different mass Kaluza-Klein (KK) partners of
the same spin.

Since different models of new physics predict different spins for
the newly discovered states, the determination of spins will be of
extreme importance for establishing the new theory. Generically,
SUSY and UED also have a different mass structure, but
model-independent measurements of both masses and spins will be
required in order to get handle on the underlying model. In this
Letter we consider the possibility of distinguishing models with the
same mass structure but the spin structure of either SUSY or
UED. We will not refer here to any particular UED model, assuming
only that it inherits all the properties (masses and couplings) of
the analyzed SUSY scenario apart from the spin. Whilst we are only
interested in the generic spin structure, we will
refer to the same-spin partners as KK-particles for ease of
notation. As a benchmark we choose two mSUGRA derived scenarios.

So far there have been a number of features suggested that would
hint at the particular spin structure at the LHC (see also~\cite{Wang:2008sw}):
\begin{itemize}
\item The total cross section~\cite{Datta:2005vx}.
\item Observation of higher KK modes~\cite{Datta:2005zs}.
\item Kinematic distributions of quarks from quark partners
decays~\cite{Nojiri:2011qn,Hallenbeck:2008hf}.
\item Particle production in vector boson fusion~\cite{Buckley:2010jv}.
\item Invariant masses of lepton-jet~\cite{Barr:2004ze,Smillie:2005ar,Athanasiou:2006ef,Wang:2006hk,Burns:2008cp} and lepton-photon~\cite{Ehrenfeld:2009rt} pairs in squark/KK-quark and gluino/KK-gluon~\cite{Alves:2006df} decay chains.
\item Kinematic reconstruction of missing
momentum~\cite{Cho:2008tj,Cheng:2010yy}.
\item Angular distributions of leptons from sleptons~\cite{Barr:2005dz,Horton:2010bg,Chen:2010ek} or $s$-channel resonances decays~\cite{Diener:2009ee}, and
$b$-jets from bottom squarks decays~\cite{Alves:2007xt}.
\end{itemize}
Most of these methods 
require significant statistics and consequently
a high luminosity (typically $\mathcal{L} \sim \mathcal{O}(100)
\fb^{-1}$) or a very specific decay chain. Here, we propose an extension of the
method originally proposed in~\cite{Barr:2005dz} for sleptons, to
the first and second generation squarks/KK-quarks. In a large class
of models, strongly-interacting states will provide the first
observation of any new physics. Therefore, they offer the
opportunity to get hints of the spin structure with early data.
Here, we consider an integrated luminosity of $\mathcal{L} = 1
\fb^{-1}$ at 14~TeV center-of-mass energy (cms), however, the method
is also applicable at lower center-of-mass energies.

\begin{figure*}[ht!]
\begin{center}
\includegraphics[scale=0.6]{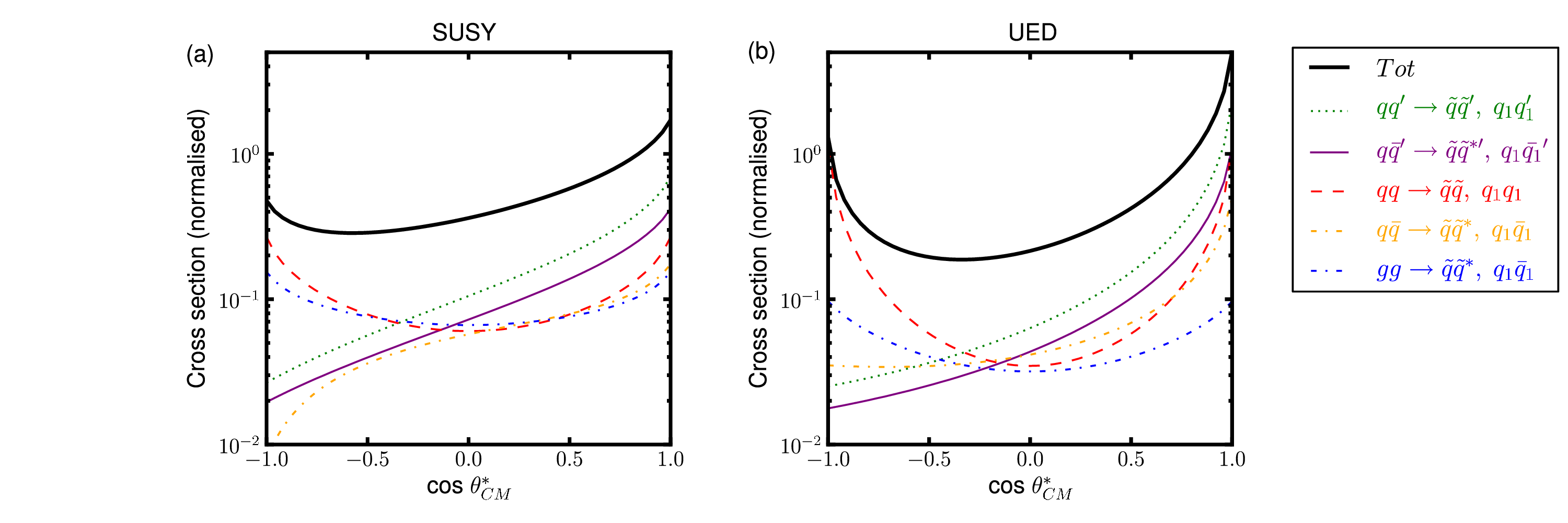}\\ \vspace{0.8cm}
\includegraphics[scale=0.6]{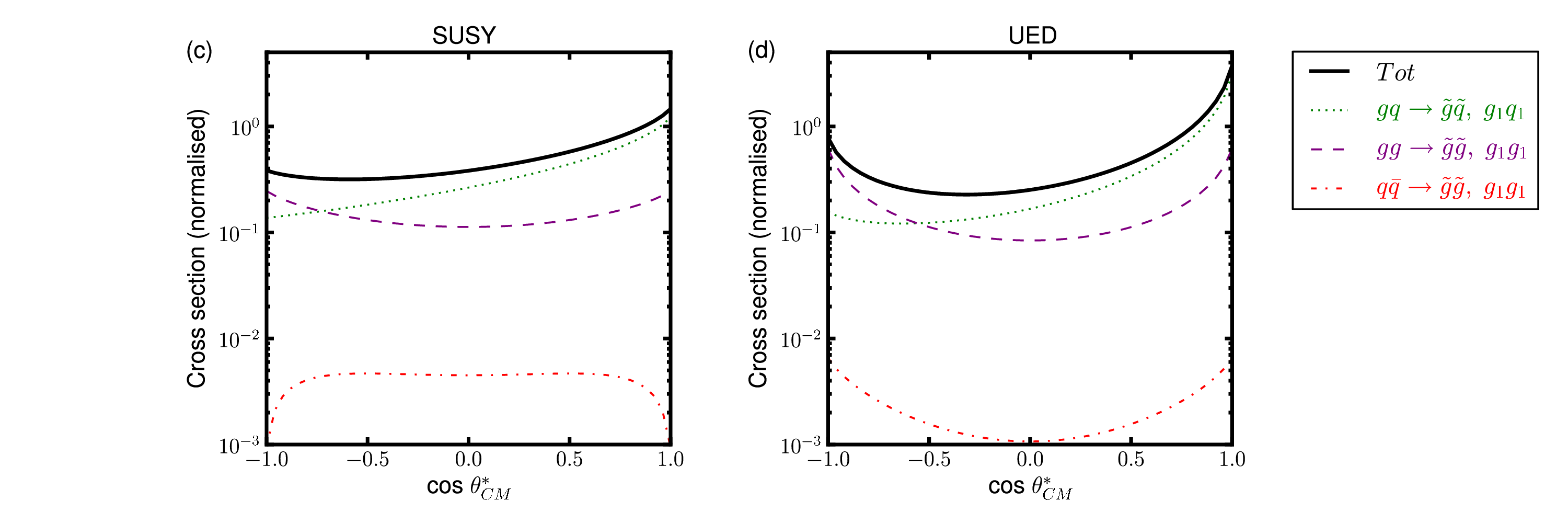}
\end{center} \vspace*{-0.4cm}
\caption{Polar distributions of (a) squarks, (b) KK-quarks,
Eq.~\eqref{eq:sqprod}; (c) gluinos, (d) KK-gluons,
Eq.~\eqref{eq:gluprod}, in the hard process CM frame normalized to
the respective total cross sections at $\sqrt{s}=14 \tev$. Contributions from left and
right states have been summed over. For simplicity we take
$m_{\tilde{q}} = m_{\tilde{g}} = 500 \gev$, however, a departure
from this assumption does not change the qualitative behaviour of
the cross sections. Note that, as dictated by the hard process
matrix element, the curves with distinguishable particles in the
initial and final state are not symmetric with respect to the
incoming parton but will be symmetric with respect to the proton
beam. \label{fig:polar} }
\end{figure*}

The Letter is organized as follows. In the next section we define
our observable and present a proof of concept that it might be
useful in studying spin structure of the underlying model. In
Section~\ref{sec:simualtion} we discuss our benchmark scenario and
details of the event simulation. Section~\ref{sec:results} contains the
results of the simulations and discussion. Finally, we conclude in
Section~\ref{sec:conclusions}.

\section{Spin-sensitive observable}
\label{sec:observable}

The production of gluinos and squarks of the 1st and 2nd generation
is in many scenarios a dominant source of supersymmetric particles.
It is, therefore, one of the most promising channels for SUSY
searches at the LHC~\cite{Aad:2009wy}. We will show in this Letter
that with early data (1~fb$^{-1}$) we can already deduce important
hints about the spin of the produced particles. We focus here on the
di-jet channel (i.e.\ at least two hard jets, see
Sec.~\ref{sec:results} for details) for which the sample sources at
the parton level are
\begin{align}
&pp \to \tilde{q}_i\, \tilde{q}_j^{(\prime)}\;,&  &pp \to
\tilde{q}_i\, \tilde{q}_j^{*(\prime)}\;,\label{eq:sqprod}\\
&pp \to \tilde{g}\, \tilde{g}\;,&  &pp \to \tilde{g}\,
\tilde{q}_i\;,\label{eq:gluprod}
\end{align}
as well as the charge conjugated processes (for squarks), followed
by the decays e.g.\
\begin{equation}\label{eq:decays}
\tilde{q}_i \to q\, \tilde{\chi}_n^0, \qquad \tilde{q}_i \to q'
\tilde{\chi}_k^\pm, \qquad \tilde{g} \to q\, \tilde{q}_i, \qquad
\mathrm{etc.},
\end{equation}
where $i,j=L,R$, $n=1,\ldots,4$, $k=1,2$ and $q'$ denotes a quark of
different flavour. Of course, the decay chains will very much depend
on the details of the spectrum of the model. One particular example
leading to a di-jet signal common in mSUGRA scenarios is
\begin{equation}\label{eq:rightsq}
pp \to \tilde{q}_R\, \tilde{q}_R \to q\, q\, \tilde{\chi}_1^0\,
\tilde{\chi}_1^0\;,
\end{equation}
whereas in UED the respective process is
\begin{equation}\label{eq:rightKK}
pp \to q_{R1}\, {q}_{R1} \to q\, q\, \gamma_1\,
\gamma_1\;,
\end{equation}
where $q_{R1}$ and $\gamma_1$ are the KK-partners of the right-handed quark and the photon, respectively.
However, many different processes from
Eqs.~(\ref{eq:sqprod})+(\ref{eq:decays}) and
(\ref{eq:gluprod})+(\ref{eq:decays}) can contribute to the di-jet
final state and therefore, we consider a fully inclusive signal. In
addition, at a hadron collider, extra QCD jets will appear when we
include initial state radiation (ISR) and final state radiation
(FSR).

Depending on the spin of the primarily produced particle, the
distributions in the center-of-mass (CM) frame of the system will be
substantially different.
Here we focus on the comparison between SUSY models and those with a
UED-like spin structure where the quark partners are spin-0 and
spin-1/2, respectively, whereas the gluon partners carry spin-1/2
and spin-1, respectively. Distributions for the production cross
section, process \eqref{eq:sqprod} and \eqref{eq:gluprod}, at
the leading order have been calculated in the literature and can be
found in
e.g.~\cite{Kane:1982hw,Harrison:1982yi,Reya:1984yz,Dawson:1983fw,Baer:1985xz,Beenakker:1996ch}
for SUSY and in~\cite{Smillie:2005ar,Macesanu:2002db} for UED. Since
the shape and relative size of the production channels depends on
the collision energy, we fold in the relevant parton density
functions (PDFs). This gives an effective polar CM angular
distribution with respect to the beam at the LHC in both cases, see
Fig.~\ref{fig:polar}. The distributions in Fig.~\ref{fig:polar} have
been calculated at the leading order, but we note that for SUSY, the
next-to-leading order (NLO) result is known~\cite{Beenakker:1996ch}.
The shape of the distributions remain nearly unchanged though and
only the total cross sections differ.

\begin{figure}
\begin{center}
\includegraphics[scale=0.6]{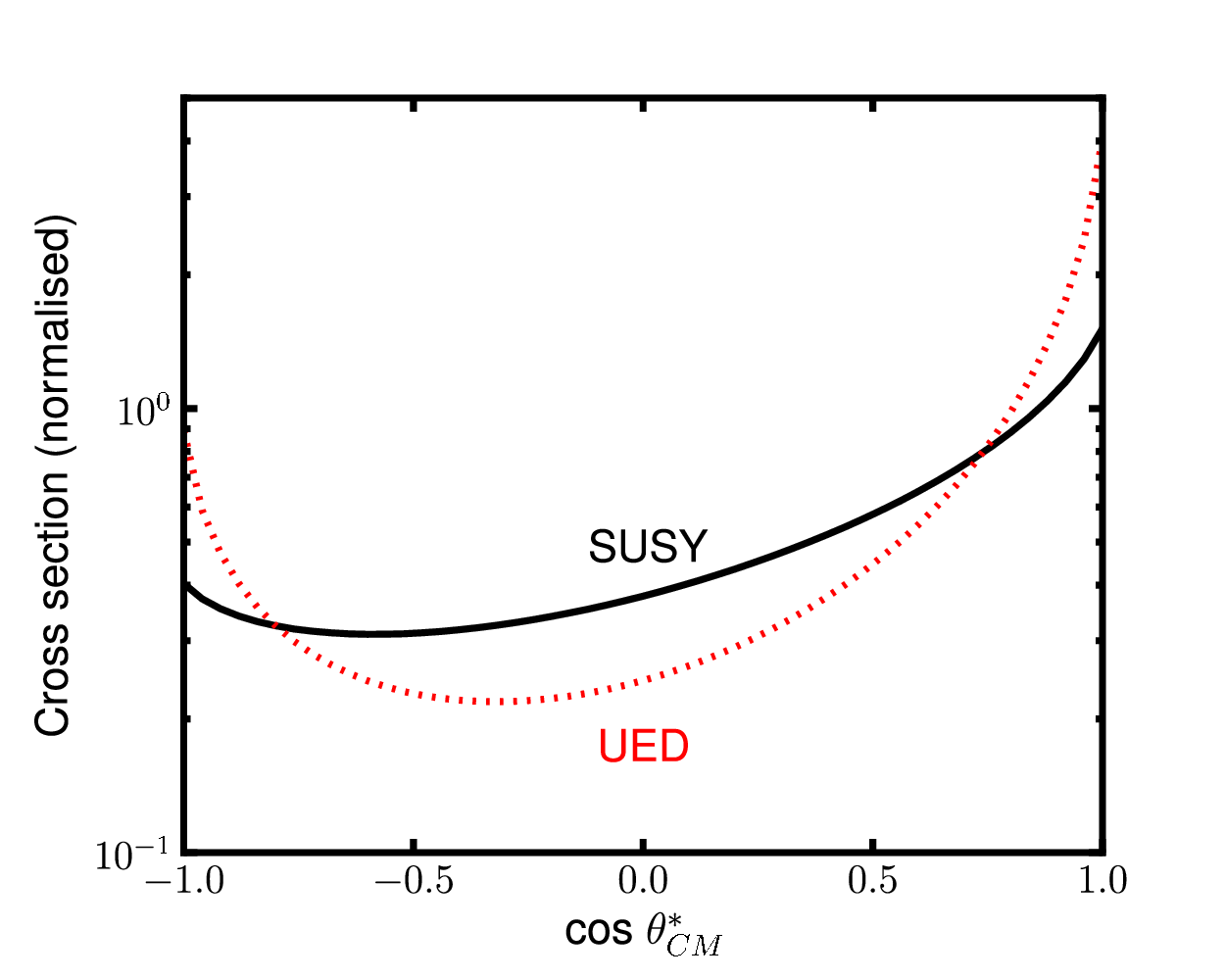}
\end{center}\vspace*{-0.4cm}
\caption{ Normalized polar distributions of squarks (black solid)
and KK-quarks (red dotted) in the hard process CM frame for
$m_{\tilde{q}} = m_{\tilde{g}} = 500 \gev$  with cms 14~TeV and with all
contributions from strong-interacting states summed
over.\label{fig:sum}}
\end{figure}

Different channels of squark/gluino (KK-quark/KK-gluon) production
exhibit a different angular dependence (see
e.g.~\cite{Baer:1985xz}), however when comparing SUSY and UED one
clearly sees that the production of supersymmetric particles (spin-0
quarks and spin-1/2 gluinos) tends to be more central compared to
KK-particles (spin-1/2 KK-quarks, spin-1 KK-gluons),
Fig.~\ref{fig:polar}. The difference between the two spin structures
is shown clearly in Fig.~\ref{fig:sum} by summing over all channels.
The main contribution to this effect comes from the fact that,
depending on the final states under consideration, SUSY particles
will be produced in either an $S$- or $P$-wave excitation. Particles
produced in a $P$-wave, due to angular momentum conservation, will
have the polar distribution vanishing for $\cos\theta^*_{CM} = \pm
1$. On the other hand, KK-particles will be dominantly produced in
$S$-waves where the distribution has a minimum at $\cos\theta^*_{CM}
= 0$, for more details see~\cite{Freitas:2001zh,Choi:2006mr}. This
conclusion is independent of the particle masses as the
KK-particles are produced in a more forward direction for all
channels discussed. In addition, the produced particles will be
heavily boosted and hence this will affect any decay products
originating from the initial particle.

\begin{figure}
\begin{center}
\includegraphics[scale=0.6]{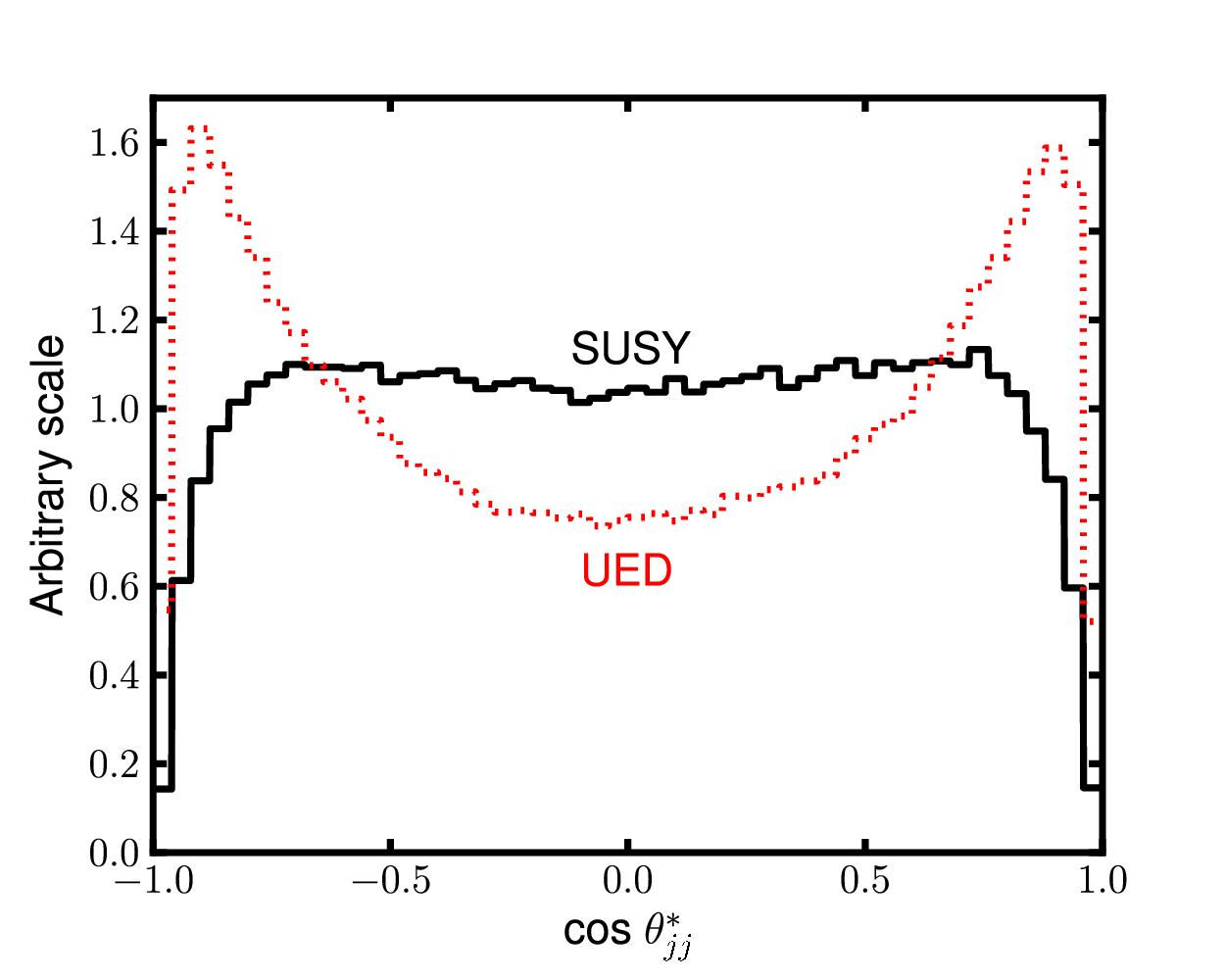}
\end{center} \vspace*{-0.4cm}
\caption{Parton level distribution of $\cos\theta_{qq}^*$,
Eq.~\eqref{eq:partonobs}, for SUSY (black solid),
Eq.~\eqref{eq:rightsq}, and UED (red dotted), Eq.~\eqref{eq:rightKK}, in the $pp$ CM frame for $m_{\tilde{q}} = m_{\tilde{g}} = 500
\gev$ and $m_{\tilde{\chi}_1^0} = 100\gev$ at $\sqrt{s}=14\tev$.
\label{fig:parton} }
\end{figure}

To probe the production distribution more directly, we propose the
following observable, originally suggested for slepton
production~\cite{Barr:2005dz}, but adapted here for squarks,
\begin{equation}\label{eq:partonobs}
\cos\theta_{qq}^* = \tanh\left(\frac{\Delta \eta_{qq}}{2}\right)
\;, \qquad \Delta\eta_{qq}= \eta_{q_1} - \eta_{q_2}\;,
\end{equation}
where $\Delta  \eta_{qq}$ is the difference of the pseudorapidities
between the two final state quarks from squarks or KK-quarks decay chains, e.g.\ Eq.~\eqref{eq:rightsq} and \eqref{eq:rightKK}, Fig.~\ref{fig:parton}. As
discussed in~\cite{Barr:2005dz} this variable is the cosine of the
polar angle of quarks with respect to the beam axis in the frame
where the pseudorapidities of the quarks are equal and opposite.
Being a function of the difference of pseudorapidities, it is
longitudinally boost invariant. This observable has proven useful in
studying the spin of sleptons~\cite{Barr:2005dz} and
sbottoms~\cite{Alves:2007xt}. Here we apply a similar approach to
the case of the 1st and 2nd generation squarks.

The distribution of the quarks in the laboratory frame is also
affected by the decay process. In the case of sparticle decay, the
angular distribution of quarks will be flat as we have a decaying
scalar,
\begin{equation}\label{eq:susyang}
\frac{d\sigma}{d \cos\theta^*_q} \propto 1\;,
\end{equation}
where $\cos\theta^*_q$ is the quark polar angle with respect to the
squark momentum, boosted to the squark rest frame. The situation is more
complicated for KK-fermions. The distribution will now depend on
both the chiral coupling structure and the polarization of the
KK-quark. For a polarized fermion we have
\begin{equation}\label{eq:uedang}
\frac{d\sigma}{d \cos\theta^*_q} \propto 1 \pm \frac{f_L^2 -
f_R^2}{f_L^2 + f_R^2} \cos\theta^*_q\;,
\end{equation}
where $f_{L,R}$ are left and right-chiral electroweak couplings,
respectively. For $f_L = f_R$ we retain the distribution from
Eq.~\eqref{eq:susyang} and there is no distinction from a scalar
decay. For the other extreme case, $f_L=0$ or $f_R=0$, the
distribution will have a triangular shape. However, since fermions
(KK-quarks in this case) will not be fully polarized, in general the
net distribution of quarks will be $\propto 1\pm \alpha
\cos\theta^*_q$ with $0<\alpha <1$. Another issue is that
anti-fermions will have an oppositely shaped distribution, cf.\
$\pm$ sign in Eq.~\eqref{eq:uedang}.

Even if the KK-quarks have a high degree of polarization, it should
be noted that at the LHC, the particles are produced with a
significant boost. Therefore the dominant structure of the
distributions will be due to the production angle of the initially
produced particles. Any angular structure from the decaying
particles can only be expected to change the distribution by a small
amount \textit{regardless of the coupling structure}. In contrast,
the majority of methods so far proposed for spin determination at
the LHC rely on a specific electroweak coupling structure in the
model (namely $f_L \gg f_R$ or $f_L \ll f_R$) \cite{Burns:2008cp}.
They would also require correct identification of particles in the
specific decay chains.

\section{Simulation and benchmark points}
\label{sec:simualtion}

In order to realistically assess the observability at the LHC, we
perform a hadronic level simulation using \texttt{Her\-wig++
2.4.2}~\cite{Bahr:2008pv,Bahr:2008tf,Gigg:2007cr} with \texttt{MRST
2004LO} PDFs~\cite{Martin:2007bv}. Events are then analyzed using
\texttt{Rivet}~\cite{Buckley:2010ar,Waugh:2006ip} and the anti-$k_t$ jet
algorithm~\cite{Cacciari:2005hq,Cacciari:2008gp} with $\Delta R =
0.5$. The jets are required to be in the central part of the
detector, i.e.\ $|\eta|<2.5$.

We include the following SM backgrounds in the analysis:
$W^\pm+$jets, $Z+$jets and $t\bar{t}$. Matrix elements are generated
using \texttt{MadGraph}~\cite{Alwall:2007st} and hadronized using
\texttt{Herwig++}. We do not include the QCD\footnote{QCD background
is subject to a large theoretical uncertainty $\mathcal{O}(100\%)$.
The correct treatment would require full detector simulation,
inclusion of higher-order effects and finally the real data. If it
turns out that the QCD background is too large, one can use the
$2j+\ell$ channel, but this in general would require a higher
integrated luminosity~\cite{Aad:2009wy}. } and di-boson backgrounds
here as these are found to be subdominant~\cite{Aad:2009wy} for the
di-jet signal after the cuts listed in Sec.~\ref{sec:results}. As
benchmark scenarios we choose the ATLAS SU3 and SU6 mSUGRA parameter
points~\cite{Aad:2009wy}. The SU3 scenario features lower masses and
is slightly above the current 
ATLAS and CMS exclusion
limits~\cite{Khachatryan:2011tk,daCosta:2011hh}, with squark masses of
$\mathcal{O}(650)\gev$. For the slightly heavier SU6 scenario the
squark masses are in the range of $\mathcal{O}(850)\gev$. Masses and
parameters for both scenarios are given in Tab.~\ref{tab:masses}.
The total NLO SUSY cross sections are $22.9 \pb$ and $6.2 \pb$ for
SU3 and SU6, respectively. We simulate the samples of signal and
backgrounds corresponding to an integrated luminosity of $1
\fb^{-1}$ at $\sqrt{s}=14$~TeV for both scenarios. Due to different
spins, the total cross section for UED is larger than for SUSY
assuming the same masses (by a factor $\sim 10$ in our scenarios).
Therefore, we normalize the number of UED events to the number of
SUSY events at NLO calculated using
\texttt{Prospino}~\cite{Beenakker:1996ch} for our comparison.

\begin{table} \renewcommand{\arraystretch}{1.2}
\begin{center}
\begin{tabular}{|c||c|c|c|c|c|} \hline
 & $m_0$ & $m_{1/2}$ & $A_0$ & $\tan\beta$ & $\mathrm{sign}(\mu)$ \\
 \hline\hline
 SU3 & $100$ & $300$ & $-300$ & $6$ & + \\ \hline
 SU6 & $320$ & $375$ & $0$ & $50$ & +\\ \hline\hline
 & $m_{\tilde{g}}$ & $m_{\tilde{q}_L}$ & $m_{\tilde{q}_R}$ &
 $m_{\tilde{\chi}_1^0}$ & $m_{\tilde{\chi}_2^0} \approx
 m_{\tilde{\chi}_1^\pm}$ \\ \hline\hline
 SU3 & $718$ & $670$ & $645$ & $119$ & $222$ \\ \hline
 SU6 & $885$ & $870$ & $840$ & $152$ & $287$ \\ \hline
\end{tabular}
\caption{mSUGRA parameters and particle masses (in GeV) for SU3 and
SU6 scenarios~\cite{Aad:2009wy}. Masses were calculated using
\texttt{SPheno~2.3}~\cite{Porod:2003um}.\label{tab:masses} }
\end{center}
\end{table}

\section{Numerical results}
\label{sec:results}

In our analysis we focus on the 2-jet signal for which we require at
least two hard jets but no hard leptons  in the final state. In order to suppress the QCD background,
relatively high-$p_T$ jets are required along with a hard
$E_T^{\mathrm{miss}}$ cut. For each of the scenarios we employ the
following set of cuts~\cite{Aad:2009wy}:
\begin{itemize}
\item At least two jets with $p_{T}^{j_1} > 150 \gev$ and $p_{T}^{j_2} >
100 \gev$ for the hardest and second-hardest jet, respectively.
\item No electrons and muons with $p_T>20 \gev$ and $|\eta_\ell|
<2.5$.
\item $\Delta\phi(j_1, E_T^{\mathrm{miss}}) > 0.2$, $\Delta\phi(j_2,
E_T^{\mathrm{miss}}) > 0.2$.
\item $E_T^{\mathrm{miss}} > 0.3 M_{\mathrm{eff}}$.
\item $M_{\mathrm{eff}} > 800 \gev$,
\end{itemize}
with the effective mass $M_{\mathrm{eff}}$
defined as
\begin{equation}
 M_{\mathrm{eff}} = p_T^{j_1} + p_T^{j_2} + E_T^{\mathrm{miss}}\;.\nonumber
\end{equation}

A detailed ATLAS study~\cite{Aad:2009wy}, including a detector
simulation, showed that the above mentioned set of cuts give a good
signal-to-background ratio ($\sim2-8$) for SUSY models with
$m_{\tilde{q}} \lesssim 900 \gev$ and high statistics even at low
integrated luminosity $\sim 1 \fb^{-1}$. Therefore, in the present
study we focus on the possibility of seeing hints of the spin
structure of the new physics at $14 \tev$ LHC in the early data. However, the
same method can be applied at lower center-of-mass energies, albeit with lower statistics.

Our observable, Eq.~\eqref{eq:partonobs}, has to be correspondingly modified at the
jet level to be
\begin{equation}\label{eq:jetobs}
\cos\theta_{jj}^* = \tanh\left(\frac{\Delta{\eta}_{jj}}{2}\right)\;, \qquad \Delta\eta_{jj}= \eta_{j_1} -
\eta_{j_2}\;,
\end{equation}
where $j_1$ and $j_2$ are the hardest and 2nd hardest jets,
respectively. We are analyzing an inclusive SUSY/UED signal,
therefore all the channels from Eqs.~\eqref{eq:sqprod} and
\eqref{eq:gluprod} will contribute. However, this turns out not to
be a problem as the gluon partners production and the associated
quark-gluon partners production all possess the desired features
observed for process~\eqref{eq:rightsq} and~\eqref{eq:rightKK}, see Fig.~\ref{fig:polar}(c)
and~\ref{fig:polar}(d).

\begin{figure*}[ht]
\begin{center}
\includegraphics[scale=0.6]{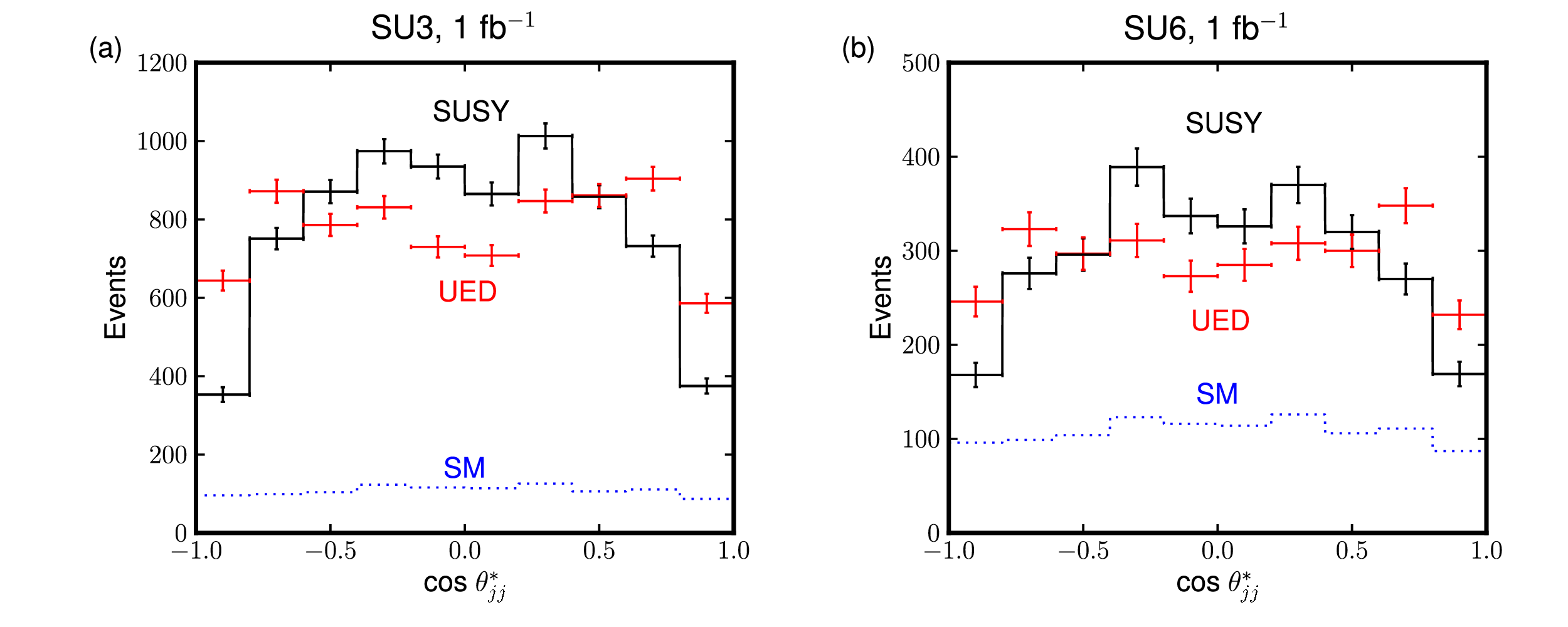}\\ \vspace{0.8cm}
\includegraphics[scale=0.6]{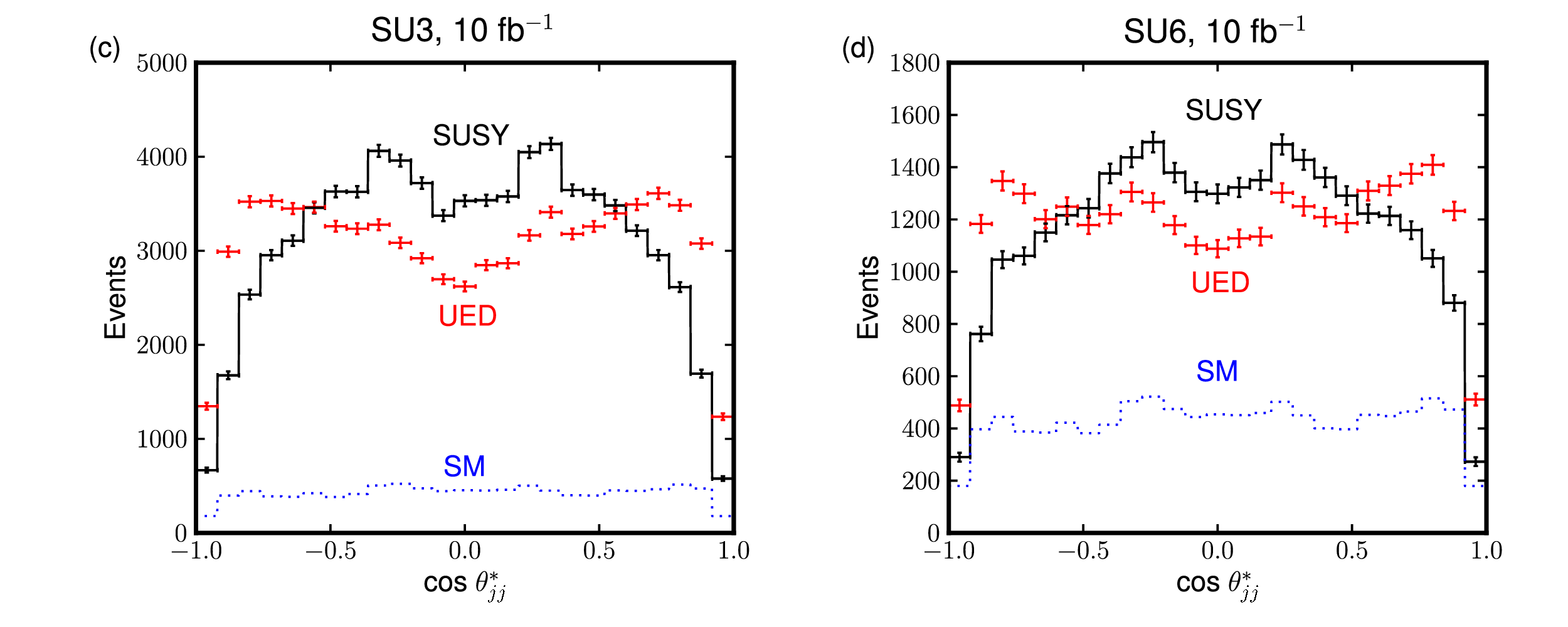}
\end{center}\vspace*{-0.4cm}
\caption{Distribution of pseudorapidity difference for jets,
Eq.~\eqref{eq:jetobs}, in SU3 ((a) and (c)) and SU6 ((b) and (d))
scenarios for SUSY (black) and UED (red) spin structure. The dotted
blue line is the SM contribution. The number of events corresponds
to an integrated luminosity $\mathcal{L} = 1 \fb^{-1}$ ((a) and (b))
and $\mathcal{L} = 10 \fb^{-1}$ ((c) and (d)) at 14~TeV cms for SUSY
particle production. The number of events for UED has been
normalized to the number of SUSY events, respectively. The
1-$\sigma$ statistical error on each point is shown.
\label{fig:etadist}}
\end{figure*}

Note that after the inclusion of hadronization and QCD radiation we are
no longer sure whether the observed hard jet originates from a
parton from the hard process. This is not a problem in the SUSY-like
scenarios discussed here because of the large mass hierarchy, but
may be a difficulty in other types of models. For example, if we take the
simplest UED model, the mass splitting between the KK-quarks and
the KK-photon is normally much smaller than in the models presented
here~\cite{Cheng:2002iz}. Consequently, the jets produced when the KK-quarks decay will
typically be relatively soft. Isolating these jets will therefore be
significantly more difficult and will certainly require a different
set of cuts than those used for this study.

\begin{figure}[t]
\begin{center}
\includegraphics[scale=0.55]{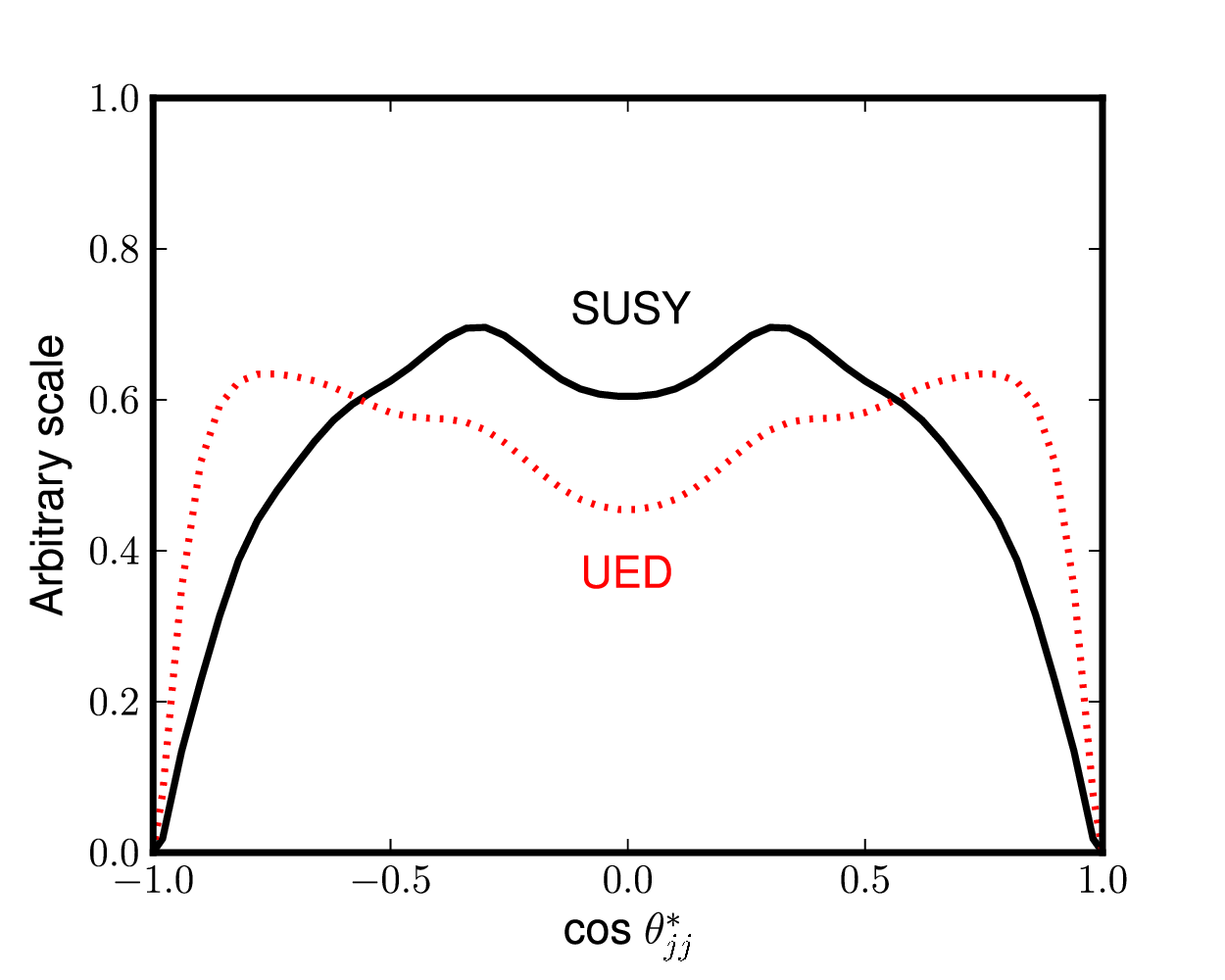}
\end{center}\vspace*{-0.4cm}
\caption{Comparison of distributions between SUSY (solid black) and
UED (dotted red) for the limiting case of very high luminosity for
the point SU3 at $\sqrt{s}= 14\tev$.
 \label{fig:etadist_high}}
\end{figure}

Figure~\ref{fig:etadist} shows that even for simulated events, large
differences remain between a model with SUSY and UED-like spin
structures for both scenarios, left and right panel, respectively.
In Fig.~\ref{fig:etadist}(a) and \ref{fig:etadist}(b) we expose the clear differences present with $1 \fb^{-1}$ of data.
Figures~\ref{fig:etadist}(c) and \ref{fig:etadist}(d) show how the situation becomes even clearer after
$10 \fb^{-1}$ have been collected. In
Fig.~\ref{fig:etadist_high} we show
the limiting case of very high statistics for SUSY and UED, where the qualitative difference between the models is clearly
visible.

Several differences can be noted when comparing the hadron-level,
Figs.~\ref{fig:etadist} and \ref{fig:etadist_high}, and the
parton-level distributions,
Fig.~\ref{fig:parton}. However, these differences can be easily
understood once the effect of the experimental cuts have been
considered. The first difference is that the number of events where
$\cos\theta^*_{jj}\sim \pm 1$ is lower than the parton-level
expectation for both SUSY and UED. This is the result
of a rapidity cut $|\eta| < 2.5$ on the final state
jets. Thus, large rapidity differences between jets are less likely
as one or both jets in the observable will not be reconstructed.

We also notice a dip in the distribution centered around
$\cos\theta^*_{jj}\sim0$.
The dip is due to the fact that the reconstructed jets have a
finite size. Two jets with a small rapidity difference can
therefore overlap if they are close in the azimuthal direction and
will not be resolved as being separate. Thus, events with this
topology will not be reconstructed.

To quantify in one number a clearly visible difference between the
distributions we introduce the following asymmetry,
\begin{align}
\mathcal{A} 
&= \frac{ N(|\cos\theta_{jj}^*|
> 0.5) - N(|\cos\theta_{jj}^*| < 0.5)}{N_{\mathrm{tot}}}\;, \label{eq:asym}
\end{align}
where $N(\ldots)$ is the number of events fulfilling the respective
condition. For our benchmark scenario SU3 at $\mathcal{L} =1 \fb^{-1}$ we obtain,
\begin{align}
\mathcal{A}^{\textrm{obs}}_{\textrm{SUSY}} &= -0.22 \pm 0.01\;, &
\mathcal{A}^{\textrm{hl}}_{\textrm{SUSY}} &= -0.226\;, \nonumber \\
\mathcal{A}^{\textrm{obs}}_{\textrm{UED}} &= \phantom{-}0.01 \pm
0.01\;, & \mathcal{A}^{\textrm{hl}}_{\textrm{UED}} &= \phantom{-} 0.016\;,\nonumber
\end{align}
where $\mathcal{A}^{\textrm{hl}}_{\textrm{SUSY}}$,
$\mathcal{A}^{\textrm{hl}}_{\textrm{UED}}$ are the high statistics limits of
the asymmetry for SUSY and UED, respectively, and
$\mathcal{A}^{\textrm{obs}}_{\textrm{SUSY}}$,
$\mathcal{A}^{\textrm{obs}}_{\textrm{UED}}$ are the observed values
with statistical errors. Thus, using the outlined method we can clearly distinguish between the two spin structures investigated.

Finally, we consider different possible mass hierarchies of squarks and gluinos (KK-quarks and KK-gluons). The original SU3 and SU6 scenarios feature similar masses of quark and gluon partners, see Tab.~\ref{tab:masses}. Here we also analyze two extreme cases, with either $m_{\tilde{q}} \ll m_{\tilde{g}}$ or  $m_{\tilde{q}} \gg m_{\tilde{g}}$. In Tab.~\ref{tab:asymmetries} we have collected the values of the asymmetry, Eq.~\eqref{eq:asym}, for different mass ratios of quark partners to gluon partners. All other parameters are the same as in the SU6 scenario. The values of the asymmetry $\mathcal{A} \approx 0$ and  $\mathcal{A} \approx -0.4$ would clearly point to a particular scenario. However, if $\mathcal{A} \approx -0.2$ we are left with an ambiguity, as this value can be obtained in both UED and SUSY scenarios. In principle, the final spin determination would require at least an approximate knowledge of the mass scales in the new physics sector.

Nevertheless, the case with heavy squarks/KK-quarks, $m_{\tilde{q}} \gg m_{\tilde{g}}$, is special since the di-jet signal is produced here by 3-body decays of gluinos/KK-gluons. In particular, neither the cuts nor the observable itself are optimized for this type of the signal. Clearly, in order to come to a decisive conclusion about the spin, an additional observable would have to be included to confirm or exclude 3-body decay case. For example, if gauginos decay to leptons a clear invariant mass, $m_{q\ell}$, edge would point to 2-body quark partner decays. On the other hand, 3-body decays of gluon partners would contain 4 partons in the final state with the similar $p_T$ distributions. In contrast, with quark partners decays, only two high-$p_T$ partons occur in the final state. Therefore, triggering on additional jet activity, together with the study of angular distributions and invariant masses, should in principle allow us to separate these two cases. We also note that heavy quarks/gluon partners would significantly lower the cross sections, hence more integrated luminosity would be required.

\begin{table}[!t]
\begin{center}
\begin{tabular}{|c||c|c|c|}\hline
     & $m_{\tilde{q}} \ll m_{\tilde{g}}$ & $m_{\tilde{q}} \approx m_{\tilde{g}}$ &
$m_{\tilde{q}} \gg m_{\tilde{g}}$ \\ \hline \hline
UED & $0.05$ & $0$ & $-0.19$ \\ \hline
SUSY & $-0.22$ & $-0.22$ & $-0.38$ \\ \hline  
\end{tabular}
\caption{Comparison of asymmetries, Eq.~\eqref{eq:asym}, in SUSY and UED for different mass scenarios of quark and gluon partners. Other parameters are the same as in the SU6 scenario. The statistical error is $0.02$, assuming the number of events as in the original SU6 scenario and an integrated luminosity $\mathcal{L} = 1 \fb^{-1}$.\label{tab:asymmetries} }
\end{center}
\end{table}

\section{Conclusions}
\label{sec:conclusions}

We have presented a method that may help to expose the spin
structure of new physics that may be seen in early LHC data. By comparing
the distributions of the di-jet rapidity distance for the models of SUSY
and UED spin structure, we found that this observable may provide
good supporting evidence to discriminate between them. The method
relies only on the spin of initially produced strongly-interacting
particles. It does not require any particular electroweak coupling
structure and it works for many possible decay chains. Therefore, it
can be applied to a wide range of scenarios. The results of the simulation show a clear distinction between models already with $\mathcal{L} = 1 \fb^{-1}$. Although we have
performed the Monte Carlo simulation for $\sqrt{s} = 14 \tev$ the
same conclusions hold at lower center-of-mass energies, albeit with lower statistics. While
more elaborate studies will be needed in order to fully confirm the spins of all individual particles in a model, early hints of the spin structure will help to direct further measurements.

\section*{Acknowledgements}
\noindent The authors wish to thank Wolfgang Ehrenfeld and Giacomo Polesello for valuable
discussions. In addition we would like to thank Frank Siegert and
Hendrik Hoeth for their help in the use of \texttt{Rivet}. JT was supported by the
Helmholtz Alliance HA-101 ``Physics at the Terascale''.
We acknowledge the support of the DFG through
the SFB (grant SFB 676/1-2006).



\bibliographystyle{utphys}
\bibliography{biblio}

\end{document}